# Genetic load makes cancer cells more sensitive to common drugs: evidence from Cancer Cell Line Encyclopedia

Ana B. Pavel*[1] and Kirill S. Korolev*[1, 2]

[1]Graduate Program in Bioinformatics, Boston University, 44 Cummington Mall, Boston, MA 02215, USA; [2]Department of Physics, Boston University, 590 Commonwealth Avenue, Boston, MA 02215, USA; e-mail: anapavel@bu.edu, korolev@bu.edu; *corresponding authors

## Abstract

Genetic alterations initiate tumors and enable the evolution of drug resistance. The pro-cancer view of mutations is however incomplete, and several studies show that mutational load can reduce tumor fitness. Given its negative effect, genetic load should make tumors more sensitive to anticancer drugs. Here, we test this hypothesis across all major types of cancer from the Cancer Cell Line Encyclopedia, that provides genetic and expression data of 496 cell lines together with their response to 24 common anticancer drugs. We found that the efficacy of 9 out of 24 drugs showed significant association with genetic load in a pan-cancer analysis. The associations for some tissue-drug combinations were remarkably strong with genetic load explaining up to 83% of the variance in the drug response. Overall, the role of genetic load depended on both the drug and the tissue type with 10 tissues being particularly vulnerable to genetic load. We also identified changes in gene expression associated with increased genetic load, which included cell-cycle checkpoints, DNA damage and apoptosis. Our results show that genetic load is an important component of tumor fitness and can predict drug sensitivity. Beyond being a biomarker, genetic load might be a new, unexplored vulnerability of cancer.

# Introduction

Cancer is triggered by the accumulation of driver alterations[1–7] and mutations conferring resistance to therapy mark the final stages of the disease. Genetic instability accelerates the appearance of adaptive mutations and is often viewed as beneficial to cancer. This view is further supported by the high prevalence of genomic instability among cancers earning it the status of a cancer hallmark[8,9].

Besides the few alterations benefiting the tumor, genetic instability produces thousands of other changes termed passengers because of their minor role in tumor progression. Traditionally, passengers have been considered as noise in cancer genomics because they obscure causative mutations that can be used as biomarkers or targets for drug design. However, several recent studies challenge this common assumption and suggest that passengers reduce tumor growth[10–18]. Each passenger could be weak and cause a relatively small reduction in fitness due to protein misfolding and aggregation, dysregulation of gene expression, or the production of neo-antigens for the immune system. The cumulative load of numerous weakly deleterious passengers could however result in a substantial fitness cost to the tumor.

Damaging mutations are known to play a major role in evolution[19–24], including the intra-host evolution of pathogens[8,25–27]. For cancer tumors, Beckman and Loeb highlighted the potential costs of genetic load more than a decade ago[16]. However, the idea of deleterious passengers has not received a lot of attention until the publication of two theoretical papers[10,11] that (i) demonstrated the plausibility of this hypothesis in the simulations of intra tumor evolution and (ii) identified a large number of protein coding changes predicted to reduce tumor fitness. Follow-up experiments in mice confirmed that passengers could severely reduce growth rates and metastatic ability of cancer cells[12]. Several clinical studies lend further support to the hypothesis of damaging passengers. Patients with highly mutated breast and ovarian cancers have been found to survive longer[28,29], and clinical trials of immunotherapies in melanoma showed efficacy only against cancers with a large number of mutations[15]. Taken together, these results suggest that genetic load reduces many components of tumor fitness[30] such as exponential growth rate, survival immune attach, and ability to metastasize. It is then natural to expect that a high load of passenger mutations should make tumors more susceptible to therapy.

Here, we test this hypothesis and investigate the relationship between genetic load and drug response across many types of cancer and anticancer drugs. This analysis has been made possible by a large-scale effort that characterized drug response in a genetically and phenotypically diverse set of 496 cell lines from the Cancer Cell Line Encyclopedia (CCLE)[31]. For each cell line, CCLE contains gene expression, copy number variations, mutations in a preselected set of genes, and growth inhibition curves for 24 anticancer drugs. These drugs included both targeted agents such as Lapatinib, which inhibits two epithelial growth factor receptors (EGFR and ERBB2)[32–34], and cytotoxic agents such as Irinotecan, a DNA Topoisomerase I inhibitor[35]. Although CCLE data has been extensively used to understand how drug response depends on specific alterations in cancer genes and changes in gene expression, the role of cumulative genetic load has not been explored previously.

Our main conclusion is that genetic load indeed makes cancer less fit, i.e. passenger mutations reduce fitness components needed to survive treatment with anticancer drugs. The efficacy of 9 out of 24 drugs in CCLE increased significantly with genetic load. For certain drugs and tissue types, the association between the drug efficacy and genetic load was especially strong, explaining up to 83% of the variance. Overall, the role of genetic load depended on both the drug and the tissue type with 10 tissues being particularly vulnerable to genetic load. The type of genetic load also mattered. In particular, copy number changes and point mutations were uncorrelated and provided independent information about the effect of genetic load on a cell line.

Genetic load also resulted in a distinct signature in gene expression changes, which included up-regulation of cell-cycle checkpoints, DNA damage, apoptosis, and other pathways. We found that over or under expression of certain growth factors, such as ERBB2 and ERBB3[33], PDGFRA, PDGFRB[37,38] and FGFR1[37,39] were strongly associated with an elevated number of point mutations. This observation further supports the findings of McFarland *et al.* [12], who experimentally demonstrated that the activation of growth factors contributes to mutagenesis. Collectively, our results confirm the important contribution of passenger alterations to tumor fitness and highlight the potential for using genetic load as a biomarker of drug response or even a therapeutic target.

## Results

Our main goal was to test whether drug sensitivity increases with genetic load. To do so, we needed to quantify genetic load and drug sensitivity, and then perform a statistical test of positive correlation. Figure 1 illustrates this approach and graphically summarizes our methods.
The drug sensitivity was quantified by the activity area, i.e. the area over the graph of relative growth inhibition vs. drug concentration. Previous studies based on CCLE data found that this metric was one of the most informative[31,40–42], so we adopted it for our analysis. Two sources of information were available to quantify the genetic load: copy number changes and point mutations. We decided to quantify them separately and then investigated possible ways to combine these measures.

*Defining genetic load*
The copy number changes were quantified by the mean length of amplifications and deletions weighted by the magnitude of the log2 change in ploidy relative to the reference genome. This measure of alteration *volume* included several aspects that are likely to influence fitness and was robust to detection errors in very short segments. For point mutations, we considered several definitions based on, for example, only synonymous, only non-sense, only missense, or all polymorphisms. The results were similar, and, here, we used the total number of variants as the simplest measure. Note that we excluded all known driver genes from the analysis not to bias our estimate of the number of passenger mutations by the inclusion of possible drivers (see Methods section for further details).

Since copy number changes and point mutations both contribute to cancer fitness, a combined measure of genetic load could have a greater predictive power. However, it is not clear *a priori* how to combine these measures at least for two reasons. First, the average fitness cost of copy number changes and point mutations could be different. Second, copy number changes and point mutations could be correlated. For example, positive correlations could appear because both types of load increase with the total number of cell divisions since cancer initiation. Negative correlations could appear due to selection for an optimal net rate at which alterations are generated[11]. In such a case, a cancer with a high rate of point mutations would have fewer copy number changes.

To address possible correlations, we computed the correlation coefficient between copy number changes and point mutation loads. The results are summarized in Figure 2, which shows no significant correlations either between raw metrics of genetic load, or between the metrics that were z-score normalized within each cancer type. In addition, we tested each tissue type independently and found no significant correlations between mutation and copy number loads (p>0.5). Given this independence between the metrics, we defined a combined genetic load as a linear combination of the copy number changes and point mutations. The weights in the linear combination were chosen to maximize its predictive ability of the drug response (see Methods). This definition also addressed the other difficulty raised above that the fitness costs of the two types of load could be different. In the following, we report the results for all three measures of genetic load: based on copy number changes only, based on point mutations only, and based on both.

*Testing for associations between genetic load and drug sensitivity*
Although testing for an association between genetic load and drug sensitivity seems straightforward, there are several caveats to consider due to cancer and drug heterogeneity. Indeed, cancer types have different amounts of copy number changes and point mutations on average and are likely to exhibit different sensitivities to the genetic load based on their phenotypic differences, for example, in protein production and metabolic rates. Drugs are also heterogeneous in their mechanism of action, and the effect of genetic load could vary greatly between the drugs. This combined variability due to drug and cancer type heterogeneity can easily obscure even a very strong association. Since many other factors such as specific mutations or expression patterns determine sensitivity to a given drug, the association between genetic load and drug response should not be particularly strong and care must be taken to control for heterogeneities in the data. Indeed, when we tested for an association between genetic load and drug response across all drugs and all cancer types we found no significant relationship for point mutation load ($\rho=0.01$, p=0.17 Spearman; r=0.003, p=0.4 Pearson) and only a weak relationship for copy number changes ($\rho=0.03$, p=0.002 Spearman; r=0.02, p=0.02 Pearson), as shown in Supplementary Figure 1.

To address the concerns raised above, we report three types of association tests: (i) pan-cancer, with all tissue types included for each drug; (ii) pan-drug, with the response effectively averaged over the drugs for each tissue type; and (iii) one for all tissue-drug combinations tested separately. The benefit of the first two approaches is that they use larger subsets of the data and reduce the number of independent tests. The benefit of the last approach is that it avoids artifacts due to cancer and drug heterogeneity and can detect effects present only for specific tissue-drug combinations.

*Pan-cancer analysis*
For each drug, we tested for a positive correlation between genetic load and drug sensitivity across all cell lines in the data set. A sample plot from this pan-cancer analysis is shown in Figure 1B and the results are summarized in Table 1. At 10% false discovery rate for Spearman correlation coefficient, we found that 9 out of 24 compounds show significant association between drug response and copy number load. In contrast, none of the associations reached significance for the point mutation load. For the combined load, we detected 7 significant associations that were identical to those for the copy number load. Table 2 shows all drugs with load-dependent activity along with their gene targets and mechanism of action.

*Pan-drug analysis*
The pan-cancer analysis demonstrated that genetic load plays a significant role in the efficacy of at least a third of drugs. Next, we examined how this association is affected by tissue heterogeneity. To this end, we compared the strength of associations between genetic load and drug response among different tissue types. To increase our statistical power, we included the data from all drugs, which affectively averages over the effects of different drugs. Figure 1B shows an example plot for this analysis and the results are summarized in Table 3.

Out of 20 tissue types in the data set, we found that 10 different tissues were associated with either the point mutation load (7 tissues) or the copy number load (4 tissues). Almost the same list of tissues were also associated with the combined load. Thus, half of the analyzed tissue types are significantly affected by the genetic load. Moreover, the correlation coefficients and statistical significance increased substantially compared to the pan-cancer analysis reaching Spearman $\rho$ as high as 0.43 and the FDR-corrected p-value as low as $10^{-8}$ (the associations based on the Pearson correlation coefficient were even stronger for some tissue types).

The observed increase in the strength of the association reflects the important contribution of tissue heterogeneity, which is also evident by the variability of the inferred correlation coefficients across different tissues. This heterogeneity could reflect some important physiological differences between cancer types that require further study. For example, if fitness costs are due to protein misfolding the differences in chaperone expression and protein production rates could be important. At the very least, our analysis shows that genetic load could be an important biomarker for drug response especially in cancers of bone, thyroid, and liver.

*Analysis of specific tissue-drug combinations*
Finally, we analyzed specific tissue-drug combinations to further control for heterogeneity and see the full predictive power of genetic load. We analyzed 9 compounds and 10 tissue types that passed the FDR corrected p-value in the pan-cancer and pan-drug analyses for the mutation and copy number loads. The results are summarized in Table 4 for copy number, point mutation, and combined loads respectively. In total, 17 associations had FDR-adjusted level below 0.1 for Spearman correlation. The true number of associations could be much higher because the small size of the data set and the large number of tests might have prevented many tissue-drug combinations from reaching statistical significance. In the Supplemental Material, we provide evidence the pan-drug and pan-tissue results remain largely the same when the data on a significant tissue-drug combinations are excluded from the analysis (Supplementary Table 2). Therefore, the selection of drugs and tissues from Tables 2 and 3 in the above analysis does not affect the statistical significance.

The correlation coefficients increased even further compared to the pan-drug analysis, explaining up to 83% of the variance that demonstrates an unexpectedly large effect of genetic load on drug sensitivity. In addition, the high correlation coefficients for tissue-drug pairs compared to pan-cancer and pan-drug analysis suggests that the fitness effects of genetic load are highly specific to both cancer type and therapeutic compound. We believe that one can obtain further insights into the biology of genetic load by trying to understand the pattern of specificity identified by our analysis. From a more clinical perspective, our results demonstrate that the tissue of origin and genetic load could be highly predictive of drug efficacy; it is therefore interesting to test whether genetic load could guide the choice of therapy for a given patient.

*The impact of the genetic load on gene expression*
The results presented above unequivocally support our hypothesis that passenger alterations reduce the fitness of cancer cells. A mechanistic understanding of this effect is however lacking. In particular, none of the drugs in CCLE was designed to increase the fitness cost of genetic load. In this section, we make a first step towards understanding the biology of passenger alterations by identifying changes in gene expression that are associated with genetic load.
We selected 50 pathways for further analysis with genes involved in apoptosis, DNA damage, cell growth, cell cycle, and other processes related to cancer initiation and progression (see Methods section for further details). Then, for each selected pathway, we tested the correlation between the genetic loads and the pathway enrichment scores computed using Gene Set Variation Analysis (GSVA) for each sample[43]. Some of the key findings are discussed below and the complete list of significant associations is provided in Supplementary Tables 3 and 4 for point mutation and copy number loads, respectively.

Overall, we found more pathways associated with mutation load compared to copy number load (26 *vs.* 7 pathways), that satisfied the FDR corrected p-value threshold of 0.1 for Spearman correlation coefficient. While this difference could truly reflect a greater and more varied response to the accumulation of point mutations, it is also possible that pathways associated with the response to copy number load are less well understood and annotated.

The top positively associated pathways with the mutation load were *apoptotic cleavage of cell adhesion proteins*, *activation of ATR in response to replication stress*[44], *cell cycle checkpoints, G1S and G2M DNA damage checkpoints*. The top negatively correlated ones were *JNK, P38, MAPK*[45,46], *GPCR*[47], *and IGF*[48] *signaling pathways*. In addition, the copy number load was also positively associated with DNA damage processes, such as *double strand break repair, G1S DNA damage checkpoints, DNA repair, ATM and E2F pathways,* Therefore, an increased genetic load may indicate increased apoptosis and DNA damage, and decreased proliferation processes.
Finally, we investigated the relationship between the expression of epithelial growth factors and genetic load. We singled out epithelial growth factors for three reasons. First, they play a major role in cancer and therefore are common targets for anticancer drugs. Second, the mechanism of action for 7 out of 9 drugs that showed significant association with genetic load involves growth factor pathways (all compounds in Table 2 except for Irinotecan and Topotecan). And third, experimental studies by McFarland et al.[12] showed that activation of growth factor receptors, such as HER2/ERBB2, significantly increases the amount of accumulating copy number changes during mutagenesis.

First, we selected 106 epithelial growth factor receptors and other related genes from UniProt database (http://www.uniprot.org) that overlapped with the CCLE data. We computed the correlation between the expression values of these growth receptors with the two genetic loads. Supporting McFarland's observations, we found significant positive associations between the epidermal growth factor receptors ERBB2 and ERBB3 and the point mutation load (FDR<0.1). In total, we found that the expression of 6 growth factor receptors and related genes was significantly positively associated with the point mutation load. In addition, the expression of 12 such genes was significantly negatively associated with the point mutation load; see Supplementary Table 5. Interestingly, BRCA1 tumor suppressor, known to be involved in EGFR regulation[49], was significantly positively associated with both the point mutation and the copy number loads. Except for a weak association with BRCA1 ($\rho=0.13$, FDR=0.07), the copy number load is not significantly associated with the expression of growth factor receptors. Although the overall picture of associations is complex, it clearly indicates that epithelial growth factor may play a role in mutation accumulation in addition to simply promoting cell growth.

## Discussion

Cancer tumor is an instance of somatic evolution that favors uncontrolled proliferation over the wellbeing of the host. To understand and control cancer, we need to understand how evolution enables and constrains tumor progression. Evolution is rarely as simple as "the survival of the fittest", and many evolutionary parameters such as mutation rates can both promote and inhibit adaptation[10,11,24,50–52]. At low mutation rates, extra mutations accelerate evolution by providing beneficial mutations and genetic diversity, which can become useful when tumor environment changes. In contrast, extra mutations could be a serious burden at high mutation rates because natural selection may fail to eliminate deleterious mutations before new ones appear. As a result, damaging mutations accumulate, reduce fitness, and interfere with the acquisition of beneficial mutations[10,11,24,50]. Where is cancer on this continuum from useful to harmful mutations?

It has been well-established that high mutation rates facilitate tumor initiation and lead to a large number of passenger alterations in cancer genomes[8]. Here, we asked whether this accumulated genetic load significantly affects cancer fitness. Using the data from CCLE, we found that both point mutations and copy number changes make cancer more vulnerable to several drugs. The magnitude of this effect is highly heterogeneous and depends strongly both on the type of cancer and the drug. For some drug-tissue combinations, genetic load explains up to 83% of the variance in the drug response among the cell lines. The statistical dependence becomes weaker as more cancer and drug types are combined together, but never loses significance indicating that the fitness cost of genetic load is a very general phenomenon.

There are several limitations of our study that might influence, but are unlikely to alter our results. The data in CCLE comes not from fresh tumor samples but from cancer cell lines that spend various amount of time growing under laboratory conditions. In addition, the fitness assays were carried out *in vitro* and do not take into account pharmacodynamics and the collateral damage to normal cells. However, the very large number of different cell lines in CCLE was chosen specifically to overcome these issues. Further, recent work demonstrated that a useful biomarker of clinical drug response can be developed from cell line data[53,54]. The performance of this cell-line-based biomarker often exceeds that of traditional biomarkers based on data derived from fresh tumor samples. Recent work also shows that previously reported inconsistencies among different cell line studies[40] can be largely resolved by using more appropriate analysis methods, which we adopted in the this study[42].

Another limitation is that point mutation data is available only for a preselected set of genes and no data on epigenetic mutations are available. While including epigenetic and whole genome sequencing data may improve the analysis, we found that including all profiled mutations or focusing just on synonymous mutations lead to similar conclusions presumably because the genes chosen for sequencing are representative of the genome overall. Moreover, the data on copy number changes was not affected by this bias and showed strong correlations with drug response as well.

Beyond demonstrating that some passenger alterations are damaging to the tumor, our analysis uncovered important associations that might shed light on tumor biology. Specifically, we found that certain tissue-drug combinations are much more sensitive to genetic load than one would expect from the average effect of genetic load on that drug or tissue. Thus, isolating processes unique to these combinations may reveal how genetic load affects fitness. Our results also suggest that epithelial growth factor receptors, such as ERBB2, ERBB3, FGFR1, FGFR4 and others, may influence the rate of mutation accumulation, a finding that echoes recent experimental observations in mice[12]. In addition to growth factors, genetic load is strongly associated with several cancer pathways reflecting their involvement in either mutagenesis or response to a high genetic load. Some of these pathways are involved in DNA damage and cell cycle response, and their action could be quite similar to the commonly studied stress response to DNA damage due to a short pulse of radiation or a mutagen. However, other identified pathways could instead represent the long-term response to the costs of a genome full of many slightly damaging mutations.

More important, the changes in gene expression associated with genetic load could provide the starting point to the design of therapies based on passenger rather than driver alterations. Simulation studies showed that such therapies are more effective than current approaches in part because normal cells have minimal genetic load[10,11]. Given the large fitness effects that we observe for drugs not designed to attack passengers, it is quite possible that therapies based on genetic load could be quite potent.

In summary, we found how to quantify genetic load based on copy number and point mutation data. We then identified 9 drugs and 10 tissue types that are significantly affected by passenger alterations. Thus, the clinical decisions for these drugs and cancers could be improved by a biomarker based on genetic load. Overall, our findings confirm that passenger mutations reduce cancer fitness and identify important physiological changes associated with genetic load. Further studies on the biology of genetic load and its therapeutic potential could benefit not only cancer research, but also diseases related to aging such as Alzheimer's disease[55], where cells are known harbor many potentially deleterious mutations.

## Methods

*CCLE Data*
In this paper, we used publicly available data from the Cancer Cell Line Encyclopedia (CCLE)[31]. CCLE consortium profiled hundreds of cell lines from different cancer types. Representation of cell lines for each cancer type was mainly based on the cancer mortality in the United States[31]. For example, for cancer types with more than 7,000 deaths/year, a maximum of 60 cell lines were profiled; for the other types, 15 was the desired minimum number of cell lines. We included data profiled from 20 different tissue types, excluding those with too few cell lines to provide sufficient statistical power. Specifically, we excluded all tissue types with fewer than 8 cell lines, such as salivary gland, biliary tract and prostate. The number of cell lines available for each tissue and each data type are shown in Supplementary Table 1. Most cell lines had both copy number point mutation profiles; see Supplementary Table 1.

We used copy number estimates profiled and normalized by CCLE[31]. This data was generated on genome-wide human Affymetrix SNP Array 6.0, normalized to log2-ratios, segmented using CBS (Circular Binary Segmentation)[56], and median centered to zero in each sample [31]. Somatic variants were measured via hybrid capture exome sequencing. A number of 1651 protein-coding genes were sequenced based on their known or potential involvement in tumor biology[31]. In the paper, we focused on common polymorphisms, so the variants with allelic fraction <10% were filtered out.

We also downloaded gene expression data generated on Affymetrix Human Genome U133 Plus 2.0 arrays and normalized using RMA (Robust Multichip Average)[31,57]. CCLE gene expression data was profiled genome-wide for 675 cell lines.

Finally, we considered the publicly available drug response data for 24 anticancer drugs profiled across 496 cell lines and 20 tissue types[31] (Supplementary Table 1). CCLE generated eight-point dose–response curves for each of the 24 compounds using an automated compound-screening platform[31]. We used the drug activity area as a measure of drug response because previous studies found it most informative[31,40–42].

*Estimating genetic load from copy number and point variants*
The copy number changes were quantified by the normalized and segmented log2-ratios relative to haploid genome. These log2-ratios are positive for copy number gain and negative for copy number loss. For each cell line, we computed the mean volume of the copy number alterations, where the volume is the absolute value of the copy number change multiplied by the region's length. Thus, amplifications and deletions contributed equally to the copy number load.

The somatic variants were profiled for 1651 protein-coding genes known to be involved in tumor biology. This choice of genes is not ideal for our purpose to estimate the genetic load caused by the accumulation of passenger mutations because some of these genes could be hot spots for driver mutations. To avoid possible biases, we filtered out the 125 known oncogenes and tumor suppressors[58]. Then, from a total number of 66613 variants, we estimated the mutation genetic load of each cell line as the total number of variants in that cell line.

Next, we used correlation analysis to identify the effect of genetic load on drug sensitivity. Because of the major differences between cancer types, the copy number load, the mutation load, and the activity area, were z-score normalized for each cancer type separately by subtracting the mean and dividing by the standard deviation. Then, we computed the correlation coefficient between the genetic load scores for a relevant subset of cell lines and the activity area of the compound data. Both Spearman and Pearson correlation coefficients are reported. However, we believe that the nonparametric Spearman coefficient is more appropriate for our analysis because it is more robust against outliers[59], and the relationship between genetic load and drug activity could be nonlinear. The statistical hypothesis tested was that the Spearman correlation coefficient is greater than zero, and we corrected the p-values using the False Discovery Rate (FDR)[60].

To combine the two genetic loads, we used a generalized linear model with lasso regularization (*cv.glmnet* function in R programming language): the activity area was the response variable and the two genetic loads were the independent variables. Then, we computed the correlation between the predicted activity area and the real value. The correlation p-values were adjusted across all computed values (those for which the lasso regularization reached a solution and the coefficients corresponding to the explanatory variables were non-zero). This procedure could slightly inflate statistical significance because the same data was used to estimate the relative weights of the two loads and to test for association between the combined load and drug response. However, this was a reasonable approach given the small number of cell line for specific tissues. Moreover, our procedure enabled an easy comparison to the correlation analyses based on a single type of genetic load. Note that the combined load largely recapitulated the results from the point mutation and copy number loads suggesting that overfitting was not a major issue in the analysis.

*Associations between genetic load and gene expression*
To identify cellular pathways most the affected by the genetic load, we tested for associations between the load and gene expression enrichment scores. First, a set of 50 pathways from MSigDb (C2, Canonical Pathways) were selected based on their relevance to DNA damage/repair, cell growth and cell cycle checkpoints. We were particularly interested in those gene sets that included the words "repair", "damage", "growth", "apoptosis", "checkpoints", "angiogenesis", "autophagy", and the most known cell growth signaling pathways such as "MAPK", "PI3K/AKT", "P53 DNA damage response", "ERBB network", "NOTCH signaling", "MTOR signaling" and "TGF-beta". To reduce noise and redundant pathway information from MSigDB, we curated the pathways using pathway maps illustrated in Cell Signaling Technology Guide (Pathways and Protocols)[61]. Next, we computed the pathway enrichment scores in each sample using GSVA[43] for the selected pathways. All cell lines with available gene expression and genetic data (620 for point mutations and 674 for copy number) were included in the analysis. For each selected pathway, we computed the correlation between the enrichment scores and the genetic loads (separately for the point mutation load and the copy number load).

Furthermore, to address the McFarland's observations that activated growth factor receptors promote mutagenesis[12], we analyzed the relationship between the expression of epithelial growth factor receptors and other related genes, to the point mutation and copy number loads. We used UniProt (http://www.uniprot.org) database (by searching for "human epithelial growth factor receptor") and overlapped the resulting list of genes with the CCLE data. We computed the correlations coefficients between gene expression of 106 growth factors and other related genes, with the two genetic loads.

## Acknowledgements

KSK was supported by a start-up fund from Boston University and by a grant from the Simons Foundation #409704. We also acknowledge the Cancer Cell Line Encyclopedia for generating the data used in this paper and making it publicly available.

## Author's contribution

ABP performed the data analysis and contributed with ideas to the methodology presented in this paper. KSK conceived the study and proposed the hypothesis and methodology presented in this paper.

## Competing financial interests

The authors declare no potential conflicts of interest.

# Figures and Tables

## Figure 1

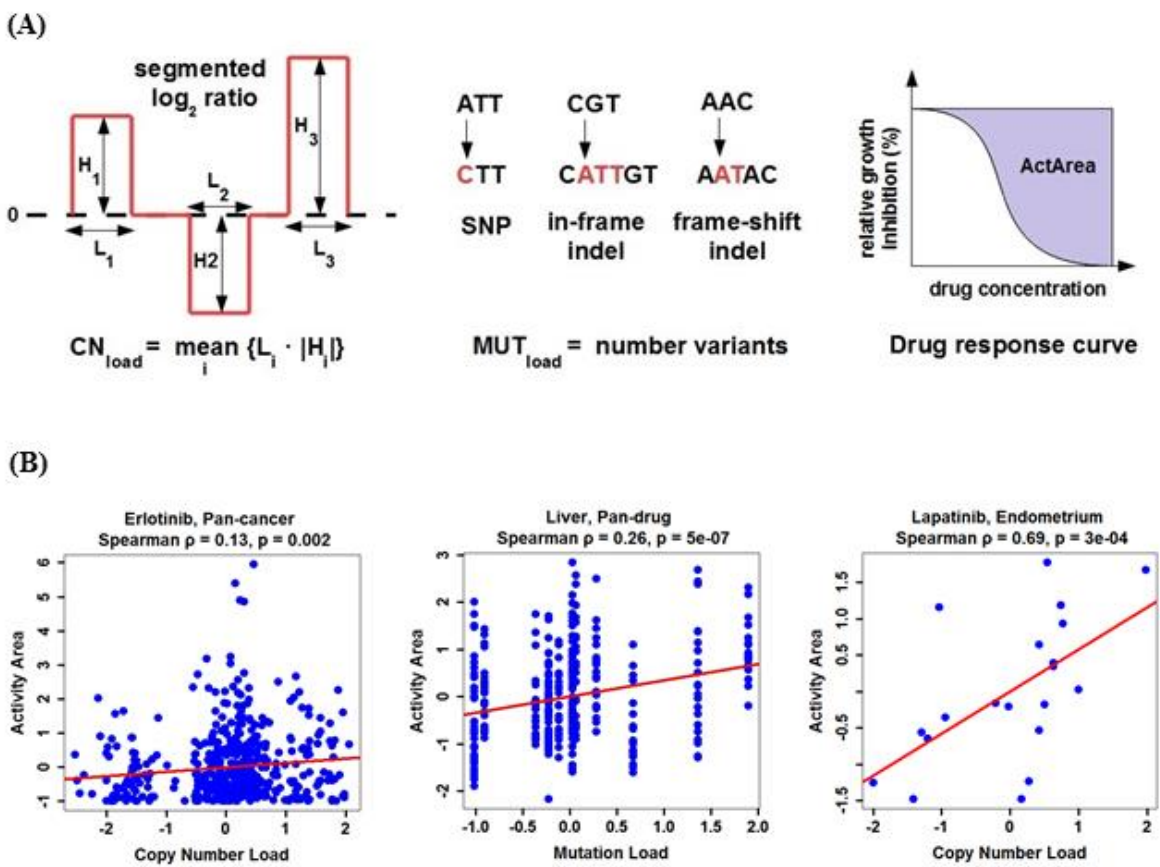

**Figure 1. Testing the relationship between genetic load and drug sensitivity.** (**A**) The first row illustrates the quantification of genetic load and drug sensitivity. From left to right: copy number load is measured as the mean absolute volume of amplifications and deletions, mutational load is defined as the total number of polymorphisms, and drug sensitivity is quantified by the area over the dose-response curve; see Methods for more details. (**B**) The second row illustrates three types of correlation analysis performed on the z-score normalized values: pan-cancer, pan-drug, and for specific tissue-drug combinations. Representative significant associations are shown. Note that the negative load reflects the normalization of genetic load accomplished by subtracting the mean load for the tissue type and then dividing by the standard deviation of the load in that tissue type.

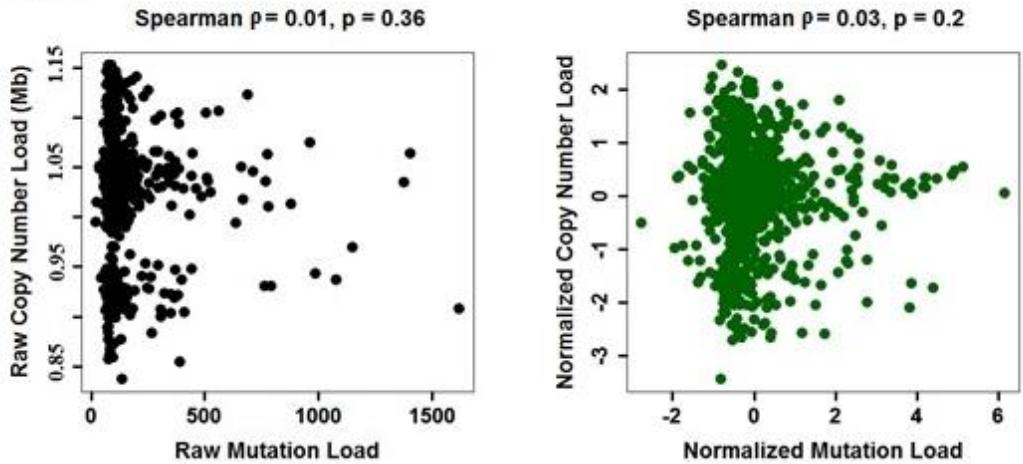

**Figure 2. Correlation between copy number and point mutation loads.** Copy number and point mutation measures of genetic load are uncorrelated. We combined all cell lines in the data set for this analysis. The left panel shows raw measures of load, while the right panel shows measures that were z-score normalized within each tissue type (i.e. divide by the the standard deviation after subtracting the mean).

**Table 1. Significant associations for the pan-cancer analysis.**

| Genetic load | Drug | Spearman | | Pearson | | No. cells |
|---|---|---|---|---|---|---|
| | | ρ | FDR | r | FDR | |
| Copy number load | Erlotinib | 0.13 | 0.04 | 0.11 | 0.04 | 486 |
| | Lapatinib | 0.12 | 0.06 | 0.11 | 0.04 | 487 |
| | TKI258 | 0.11 | 0.06 | 0.13 | 0.04 | 487 |
| | Sorafenib | 0.09 | 0.06 | 0.12 | 0.04 | 486 |
| | AEW541 | 0.1 | 0.06 | 0.09 | 0.11 | 486 |
| | Irinotecan | 0.12 | 0.06 | 0.09 | 0.21 | 304 |
| | Nilotinib | 0.1 | 0.06 | 0.07 | 0.24 | 403 |
| | TAE684 | 0.09 | 0.06 | 0.04 | 0.44 | 487 |
| | Topotecan | 0.09 | 0.06 | 0.07 | 0.21 | 487 |
| Combined load | Erlotinib | 0.13 | 0.02 | 0.11 | 0.02 | 441 |
| | Lapatinib | 0.11 | 0.03 | 0.12 | 0.02 | 442 |
| | TKI258 | 0.11 | 0.03 | 0.12 | 0.02 | 442 |
| | AEW541 | 0.1 | 0.03 | 0.08 | 0.06 | 441 |
| | Topotecan | 0.1 | 0.03 | 0.07 | 0.09 | 442 |
| | Nilotinib | 0.1 | 0.03 | 0.06 | 0.12 | 363 |
| | Sorafenib | 0.08 | 0.05 | 0.11 | 0.02 | 441 |

**Table 2. Significant drugs in pan-cancer analysis and their gene targets (the targets, predictors of sensitivity and mechanisms of action were included from the CCLE study[31]).**

| Drug | Gene targets | Predictor of sensitivity | Mechanism of action |
|---|---|---|---|
| Erlotinib | EGFR | EGFR mutation | EGFR inhibitor |
| Lapatinib | ERBB2, EGFR | ERBB2 expression | EGFR and ERBB2 inhibitor |
| TKI258 | EGFR, FGFR1, PDGFRbeta, VEGFR-1, KDR | Unknown | Multi-kinase inhibitor |
| Sorafenib | FLT3, C-KIT, PDGFRbeta, RET, Raf kinase B, Raf kinase C, VEGFR-1, KDR, FLT4 | Unknown | Multi-kinase inhibitor |
| AEW541 | IGF1R | IGF1R expression | Kinase inhibitor |
| Irinotecan | Topoisomerase I | Unknown | DNA Topoisomerase I Inhibitor |
| Nilotinib | Abl/Bcr-Abl | Unknown | Abl Inhibitor |
| TAE684 | ALK | Unknown | ALK Inhibitor |
| Topotecan | Topoisomerase I | Unknown | DNA Topoisomerase I Inhibitor |

**Table 3. Significant associations for the pan-drug analysis.**

| Genetic load | Tissue type | Spearman $\rho$ | Spearman FDR | Pearson r | Pearson FDR | No. cells |
|---|---|---|---|---|---|---|
| Point mutation load | BONE | 0.34 | $10^{-6}$ | 0.33 | $10^{-6}$ | 260 |
| | LIVER | 0.26 | $10^{-5}$ | 0.27 | $10^{-5}$ | 338 |
| | THYROID | 0.43 | $10^{-5}$ | 0.44 | $10^{-5}$ | 120 |
| | CENTRAL NERVOUS SYSTEM | 0.18 | $10^{-4}$ | 0.13 | $10^{-3}$ | 576 |
| | STOMACH | 0.15 | 0.01 | 0.31 | $10^{-7}$ | 349 |
| | PANCREAS | 0.09 | 0.05 | 0.07 | 0.16 | 599 |
| | LUNG | 0.04 | 0.08 | 0.04 | 0.09 | 1997 |
| Copy number load | SKIN | 0.16 | $10^{-5}$ | 0.15 | $10^{-5}$ | 936 |
| | LIVER | 0.14 | 0.01 | 0.13 | 0.02 | 434 |
| | HAEMATOPOIETIC AND LYMPHOID TISSUE | 0.07 | 0.01 | 0.06 | 0.03 | 1677 |
| | ENDOMETRIUM | 0.15 | 0.01 | 0.08 | 0.14 | 458 |
| Combined load | BONE | 0.38 | $10^{-8}$ | 0.34 | $10^{-8}$ | 236 |
| | LIVER | 0.29 | $10^{-7}$ | 0.37 | $10^{-11}$ | 338 |
| | THYROID | 0.43 | $10^{-6}$ | 0.44 | $10^{-6}$ | 120 |
| | CENTRAL NERVOUS SYSTEM | 0.18 | $10^{-5}$ | 0.13 | $10^{-3}$ | 576 |
| | SKIN | 0.13 | $10^{-4}$ | 0.13 | $10^{-4}$ | 841 |
| | ENDOMETRIUM | 0.15 | $10^{-3}$ | 0.08 | 0.05 | 458 |
| | HAEMATOPOIETIC AND LYMPHOID TISSUE | 0.08 | $10^{-3}$ | 0.07 | $10^{-3}$ | 1535 |
| | STOMACH | 0.15 | $10^{-3}$ | 0.31 | $10^{-8}$ | 349 |
| | PANCREAS | 0.10 | 0.01 | 0.07 | 0.05 | 551 |
| | LARGE INTESTINE | 0.09 | 0.03 | 0.12 | 0.01 | 488 |

**Table 4. Significant associations for tissue-drug combinations.**

| Genetic load | Tissue type | Drug | Spearman | | Pearson | | No. cells |
|---|---|---|---|---|---|---|---|
| | | | ρ | FDR | r | FDR | |
| Point mutation load | CENTRAL NERVOUS SYSTEM | Sorafenib | 0.69 | 0.01 | 0.56 | 0.18 | 25 |
| Copy number load | ENDOMETRIUM | Lapatinib | 0.69 | 0.03 | 0.58 | 0.11 | 20 |
| | LIVER | Irinotecan | 0.83 | 0.06 | 0.79 | 0.11 | 11 |
| | STOMACH | TAE684 | 0.64 | 0.06 | 0.59 | 0.11 | 19 |
| Combined load | CENTRAL NERVOUS SYSTEM | Sorafenib | 0.69 | $10^{-3}$ | 0.56 | 0.02 | 25 |
| | ENDOMETRIUM | Lapatinib | 0.69 | $10^{-3}$ | 0.58 | 0.02 | 20 |
| | LIVER | Lapatinib | 0.6 | 0.04 | 0.68 | 0.02 | 15 |
| | STOMACH | TAE684 | 0.65 | 0.04 | 0.60 | 0.04 | 15 |
| | ENDOMETRIUM | Erlotinib | 0.52 | 0.04 | 0.47 | 0.04 | 20 |
| | BONE | Sorafenib | 0.75 | 0.04 | 0.59 | 0.06 | 10 |
| | BONE | Nilotinib | 0.81 | 0.04 | 0.59 | 0.08 | 8 |
| | LIVER | Nilotinib | 0.73 | 0.04 | 0.84 | 0.02 | 9 |
| | SKIN | Erlotinib | 0.36 | 0.04 | 0.33 | 0.05 | 36 |
| | BONE | Erlotinib | 0.7 | 0.04 | 0.54 | 0.08 | 10 |
| | PANCREAS | Irinotecan | 0.52 | 0.05 | 0.53 | 0.04 | 16 |
| | STOMACH | TKI258 | 0.51 | 0.05 | 0.48 | 0.06 | 15 |
| | SKIN | Topotecan | 0.33 | 0.05 | 0.37 | 0.04 | 36 |
| | LIVER | Irinotecan | 0.71 | 0.08 | 0.81 | 0.04 | 7 |
| | BONE | AEW541 | 0.58 | 0.08 | 0.67 | 0.04 | 10 |
| | HAEMATOPOIETIC AND LYMPHOID TISSUE | TKI258 | 0.21 | 0.08 | 0.29 | 0.04 | 65 |
| | HAEMATOPOIETIC AND LYMPHOID TISSUE | Erlotinib | 0.2 | 0.08 | 0.19 | 0.08 | 65 |

# Supplementary Figures and Tables

## Supplementary Figure 1

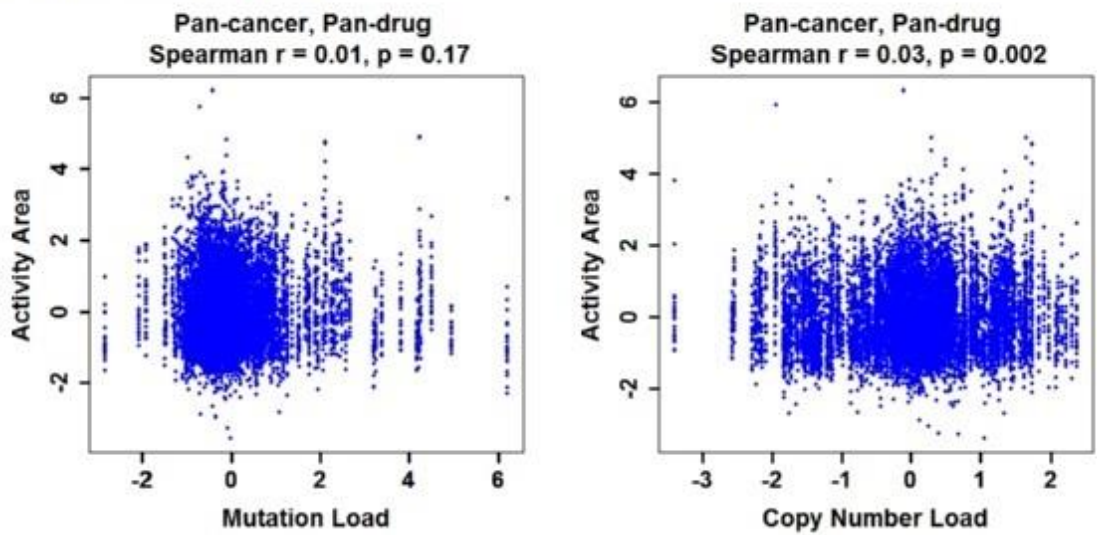

**Supplementary Figure 1**. Pan-data analysis for the mutation load (left) and the copy number load (right).

**Supplementary Table 1. Twenty CCLE tissue types analyzed in this work.**

| TISSUE TYPE | Number of cell lines with profiled copy number data (1017) | Number of cell lines with profiled point mutation data (888) |
|---|---|---|
| HAEMATOPOIETIC AND LYMPHOID TISSUE | 188 | 165 |
| LUNG | 185 | 172 |
| CENTRAL NERVOUS SYSTEM | 68 | 44 |
| SKIN | 61 | 53 |
| BREAST | 59 | 51 |
| LARGE INTESTINE | 59 | 56 |
| OVARY | 53 | 47 |
| PANCREAS | 44 | 37 |
| STOMACH | 39 | 34 |
| KIDNEY | 36 | 22 |
| UPPER AERODIGESTIVE TRACT | 31 | 31 |
| ENDOMETRIUM | 28 | 27 |
| BONE | 28 | 22 |
| LIVER | 27 | 24 |
| OESOPHAGUS | 27 | 25 |
| URINARY TRACT | 24 | 24 |
| SOFT TISSUE | 21 | 18 |
| AUTONOMIC GANGLIA | 17 | 15 |
| THYROID | 12 | 11 |
| PLEURA | 10 | 10 |

**Supplementary Table 2. Significant associations for tissue-drug combinations with exclusion of tested drug-tissue combination from pan cancer and pan drug analyses.** In the analysis of drug-tissue combinations, we preselected drugs and tissues based on the association with genetic load. Thus, some of the data was used twice: once in the selection step and once in the correlation test. Overall, the number of cell lines used twice was typically less than 10%, and, in many cases, pre-selection was based on a different measure of genetic load than the correlation test, i.e. on a different data. Nevertheless, we modified the analysis to ensure that statistical significance is not affected. We screened all tissue-drug combinations, not just those from Tables 2 and 3. For each combination, we first tested whether the tissue and the drug are associated with genetic load when the data from the combination is excluded. The combination was considered further only if pan-drug and pan-cancer analyses (with exclusion) showed positive association with the genetic load (Spearman FDR less than 0.1). For each combination passing this test, we computed the correlation between the Activity Area and the genetic load and performed FDR calculation based on the total number of combinations passing the test for each load.

| Genetic load | Tissue type | Drug | Spearman | | Pearson | | No. cells |
|---|---|---|---|---|---|---|---|
| | | | $\rho$ | FDR | r | FDR | |
| Point mutation load | LUNG | Lapatinib | 0.39 | 0.08 | 0.55 | 0.02 | 15 |
| Copy number load | ENDOMETRIUM | Lapatinib | 0.69 | 0.01 | 0.58 | 0.04 | 20 |
| | LIVER | Irinotecan | 0.83 | 0.01 | 0.79 | 0.03 | 11 |
| | ENDOMETRIUM | Erlotinib | 0.64 | 0.06 | 0.47 | 0.12 | 20 |
| Combined load | ENDOMETRIUM | Lapatinib | 0.69 | $10^{-3}$ | 0.58 | 0.01 | 20 |
| | LIVER | Lapatinib | 0.69 | 0.01 | 0.68 | 0.01 | 15 |
| | ENDOMETRIUM | Erlotinib | 0.6 | 0.01 | 0.47 | 0.03 | 20 |
| | LIVER | Nilotinib | 0.65 | 0.02 | 0.84 | 0.01 | 9 |
| | STOMACH | TKI258 | 0.52 | 0.03 | 0.48 | 0.05 | 15 |
| | LIVER | Topotecan | 0.75 | 0.08 | 0.57 | 0.03 | 15 |

**Supplementary Table 3. Associations between cancer pathways and point mutation load.**

| PATHWAY | Spearman | | Pearson | |
|---|---|---|---|---|
| | ρ | FDR | r | FDR |
| REACTOME_APOPTOTIC_CLEAVAGE_OF_CELL_ADHESION_PROTEINS | 0.25 | $10^{-7}$ | 0.27 | $10^{-9}$ |
| ST_JNK_MAPK_PATHWAY | -0.24 | $10^{-7}$ | -0.13 | 0.01 |
| REACTOME_GPCR_DOWNSTREAM_SIGNALING | -0.23 | $10^{-6}$ | -0.26 | $10^{-8}$ |
| PID_P38_ALPHA_BETA_PATHWAY | -0.21 | $10^{-5}$ | -0.15 | $10^{-3}$ |
| KEGG_MAPK_SIGNALING_PATHWAY | -0.21 | $10^{-5}$ | -0.12 | 0.01 |
| REACTOME_ACTIVATION_OF_ATR_IN_RESPONSE_TO_REPLICATION_STRESS | 0.19 | $10^{-4}$ | 0.07 | 0.18 |
| REACTOME_CELL_CYCLE_CHECKPOINTS | 0.19 | $10^{-4}$ | 0.11 | 0.02 |
| REACTOME_SIGNALING_BY_GPCR | -0.19 | $10^{-4}$ | -0.26 | $10^{-9}$ |
| REACTOME_REGULATION_OF_INSULIN_LIKE_GROWTH_FACTOR_IGF_ACTIVITY_BY_INSULIN_LIKE_GROWTH_FACTOR_BINDING_PROTEINS_IGFBPS | -0.19 | $10^{-4}$ | -0.13 | 0.01 |
| REACTOME_G2_M_CHECKPOINTS | 0.18 | $10^{-4}$ | 0.07 | 0.18 |
| PID_LYMPH_ANGIOGENESIS_PATHWAY | -0.15 | $10^{-3}$ | -0.17 | $10^{-4}$ |
| REACTOME_G2_M_DNA_DAMAGE_CHECKPOINT | 0.13 | 0.01 | 0.05 | 0.38 |
| REACTOME_P53_DEPENDENT_G1_DNA_DAMAGE_RESPONSE | 0.12 | 0.01 | 0.12 | 0.01 |
| REACTOME_P53_INDEPENDENT_G1_S_DNA_DAMAGE_CHECKPOINT | 0.12 | 0.01 | 0.05 | 0.34 |
| PID_E2F_PATHWAY | 0.12 | 0.01 | 0.06 | 0.23 |
| REACTOME_GROWTH_HORMONE_RECEPTOR_SIGNALING | -0.11 | 0.02 | -0.07 | 0.21 |
| KEGG_MTOR_SIGNALING_PATHWAY | -0.11 | 0.02 | -0.14 | $10^{-3}$ |
| REACTOME_DNA_REPAIR | 0.11 | 0.02 | 0.03 | 0.61 |
| REACTOME_APOPTOTIC_EXECUTION_PHASE | 0.11 | 0.02 | 0.17 | $10^{-4}$ |
| REACTOME_EXTRINSIC_PATHWAY_FOR_APOPTOSIS | -0.10 | 0.03 | -0.02 | 0.71 |
| PID_ERBB_NETWORK_PATHWAY | 0.10 | 0.03 | 0.12 | 0.01 |
| PID_ATM_PATHWAY | 0.10 | 0.04 | 0.02 | 0.71 |
| REACTOME_APOPTOTIC_CLEAVAGE_OF_CELLULAR_PROTEINS | 0.09 | 0.04 | 0.12 | 0.01 |
| KEGG_NOTCH_SIGNALING_PATHWAY | 0.09 | 0.05 | 0.15 | $10^{-3}$ |
| REACTOME_NRAGE_SIGNALS_DEATH_THROUGH_JNK | -0.09 | 0.05 | 0.14 | $10^{-3}$ |
| REACTOME_SIGNALING_BY_HIPPO | 0.08 | 0.07 | 0.05 | 0.33 |

**Supplementary Table 4. Associations between cancer pathways and copy number load.**

| PATHWAY | Spearman | | Pearson | |
|---|---|---|---|---|
| | ρ | FDR | r | FDR |
| PID_P38_ALPHA_BETA_DOWNSTREAM_PATHWAY | 0.12 | 0.03 | 0.11 | 0.06 |
| REACTOME_HOMOLOGOUS_RECOMBINATION_REPAIR_OF_REPLICATION_INDEPENDENT_DOUBLE_STRAND_BREAKS | 0.12 | 0.03 | 0.12 | 0.06 |
| REACTOME_DOUBLE_STRAND_BREAK_REPAIR | 0.12 | 0.03 | 0.12 | 0.06 |
| PID_E2F_PATHWAY | 0.12 | 0.03 | 0.07 | 0.39 |
| REACTOME_P53_INDEPENDENT_G1_S_DNA_DAMAGE_CHECKPOINT | 0.11 | 0.04 | 0.09 | 0.14 |
| PID_ATM_PATHWAY | 0.11 | 0.04 | 0.10 | 0.07 |
| REACTOME_DNA_REPAIR | 0.10 | 0.09 | 0.09 | 0.14 |

**Supplementary Table 5. Associations between epithelial growth factor receptors and other related genes and point mutation load.**

| GENE | Spearman | | Pearson | |
|---|---|---|---|---|
| | ρ | FDR | r | FDR |
| TGFB1 | -0.21 | $10^{-5}$ | -0.14 | 0.01 |
| PML | -0.21 | $10^{-5}$ | -0.20 | $10^{-4}$ |
| ERBB3 | 0.18 | $10^{-4}$ | 0.16 | $10^{-3}$ |
| SGK3 | 0.19 | $10^{-4}$ | 0.08 | 0.16 |
| SMAD3 | -0.17 | $10^{-4}$ | -0.08 | 0.17 |
| TGFB2 | -0.17 | $10^{-4}$ | -0.11 | 0.03 |
| MMP2 | -0.17 | $10^{-4}$ | -0.12 | 0.02 |
| MYLK | -0.16 | $10^{-3}$ | -0.13 | 0.01 |
| PAK2 | -0.16 | $10^{-3}$ | -0.12 | 0.03 |
| FGFR4 | 0.15 | $10^{-3}$ | 0.17 | $10^{-3}$ |
| JAG2 | 0.14 | $10^{-3}$ | 0.14 | 0.01 |
| BRCA1 | 0.13 | 0.01 | 0.05 | 0.45 |
| FGFR1 | -0.14 | 0.01 | -0.04 | 0.60 |
| VEGFC | -0.13 | 0.01 | -0.15 | $10^{-3}$ |
| ELF4 | -0.13 | 0.01 | -0.10 | 0.07 |
| PTEN | -0.11 | 0.04 | -0.04 | 0.54 |
| ERBB2 | 0.10 | 0.06 | 0.14 | 0.01 |
| INHBA | -0.10 | 0.07 | -0.09 | 0.12 |